\documentclass{article}

\usepackage{epsfig}
\usepackage{graphics}
\usepackage{graphicx}

\usepackage{url}
\usepackage{subfigure}
\usepackage{hyperref}
\usepackage{multirow}

\title{Following the spacial dynamics of COVID-19 in Mexico and some notes}

\author{Genaro J. Mart{\'i}nez$^{1,2}$, Magali C\'ardenas Tapia$^{3}$, Ricardo Antonio \\ Tena N\'u\~nez$^{4}$, Adriana de la Paz S\'anchez Moreno$^{1}$}

\date{June 2021}

\begin{document}

\maketitle

\begin{centering}
$^1$ Artificial Life Robotics Laboratory, Escuela Superior de C\'omputo, Instituto Polit\'ecnico Nacional, M\'exico. \\
\url{gjuarezm@ipn.mx}, \url{apsanchezm@ipn.mx} \\
$^2$ Unconventional Computing Laboratory, University of the West of England, Bristol, United Kingdom. \\
$^3$ Escuela Superior de Comercio y Administraci\'on, Unidad Tepepan, Instituto Polit\'ecnico Nacional, M\'exico. \\
\url{mcardenast@ipn.mx} \\
$^4$ Escuela Superior de Ingenier{\'i}a y Arquitectura, Unidad Tecamachalco, Instituto Polit\'ecnico Nacional, M\'exico. \\
\url{rtena@ipn.mx} \\
\end{centering}

\begin{abstract}
\noindent
Actually, after one year it is recognized that the evolution of COVID-19 is different in each country or region around the world. In this paper, we do a revision to the date about COVID-19 evolution in Mexico, we explain where the main epicenter and states with most high impact. Mexico has a particular geographical position in the American continent because it is a natural bridge between the USA and Latin America, that represents a special point of propagation because between other facts this virus is transported by people of different nationalities migrating to the USA. The research in this paper helps to understand why Mexico is one of the countries with the most high mortality impact by this new virus and how the lockdown works in the population. Finally, we give a practical perspective as this evolution is a complex system.
\vspace{3mm}

\end{abstract}

\newpage

%%%%%%%%%%%%%%
\section{Introduction}

In this paper, we present a research dedicated particularly to the spatial dynamics of COVID-19 evolution in Mexico. Across this spatial analysis we can recognize the main epicenter in the Mexican republic. We discuss briefly some political, economic, cultural and social aspects to try to explain and understand why Mexico is one of the most affected countries around the world with a high number of deaths by this virus.

This way, the paper is organized as follows. Section 2 shows a general social and political description about how started COVID-19 in Mexico and its terrible handling from several sectors. So for such a description we focus our attention mainly by the number of deaths derived by COVID-19 in every state. In Section 3, we display the spatial dynamics of COVID-19 in the whole Mexican republic by the number of positive cases per day during 14 months. Finally, in Section 4 we discuss how the COVID-19 dynamics displays complex behavior and unpredictability.

%%%%%%%%%%%%%%
\section{General overview}

Mexico displays a particular case of study as the third country with more deaths by COVID-19 in the world. It is clear that the Mexican government began and continued with a very bad handling of the pandemic. Of course, we can find a lot of circumstances, actions and society participation which also do not help to get a better preparation. Despite the experience several months ago mainly from Europe and later from the USA. Mexico did not have any reaction from their boundaries, airports, ports and local transport (mobility). So, Mexico is a natural bridge of migrants traveling to the USA and Canada.

\begin{figure}[th]
\centering
\includegraphics[width=1\textwidth]{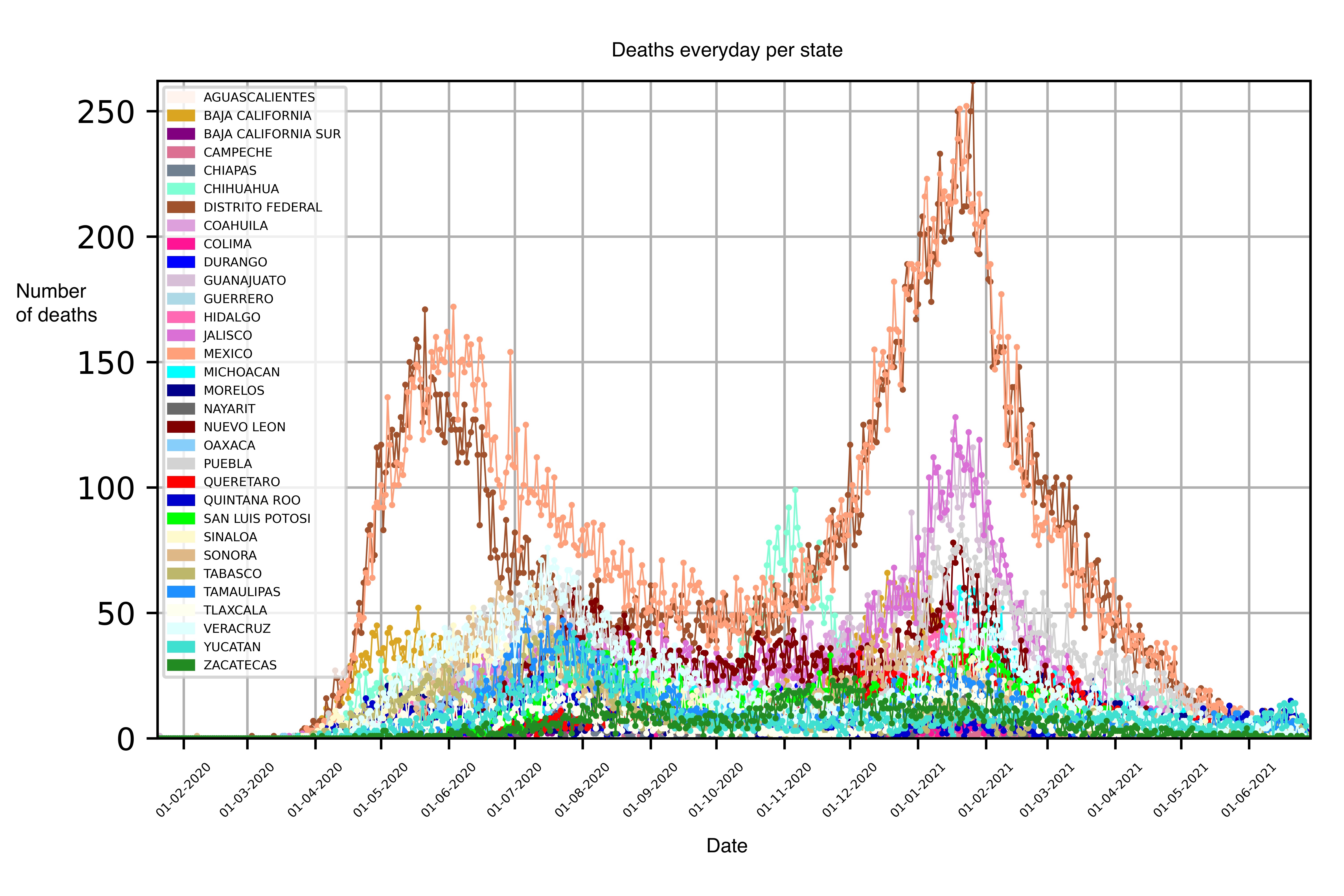}
\caption{Deaths everyday per state in Mexico from February 2019 to June 2021.}
\label{2DPlotDead32States}
\end{figure}

\begin{figure}%[th]
\begin{center}
\subfigure[]{\scalebox{0.32}{\includegraphics{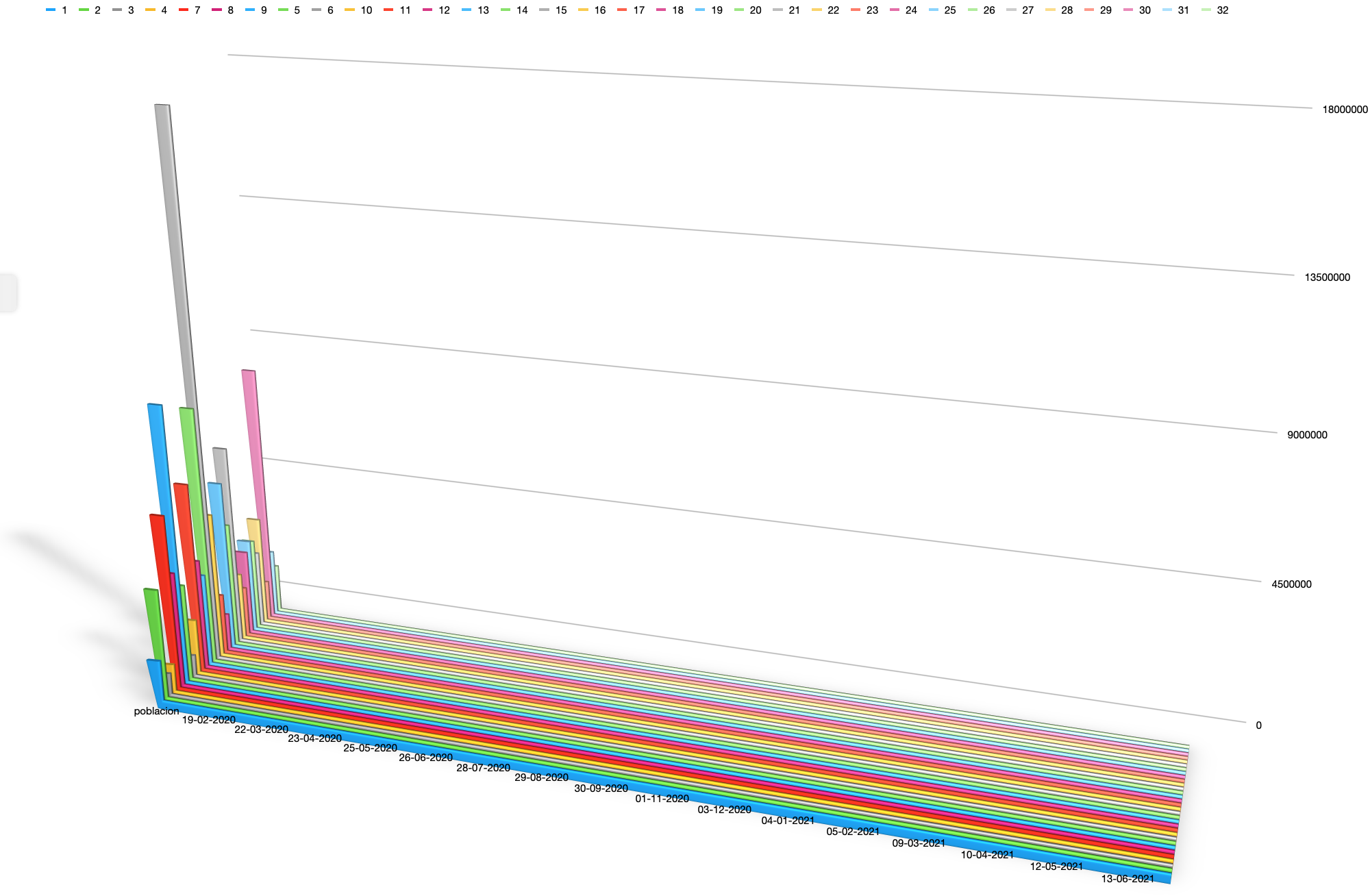}}} %\hspace{0.7cm}
\subfigure[]{\scalebox{0.32}{\includegraphics{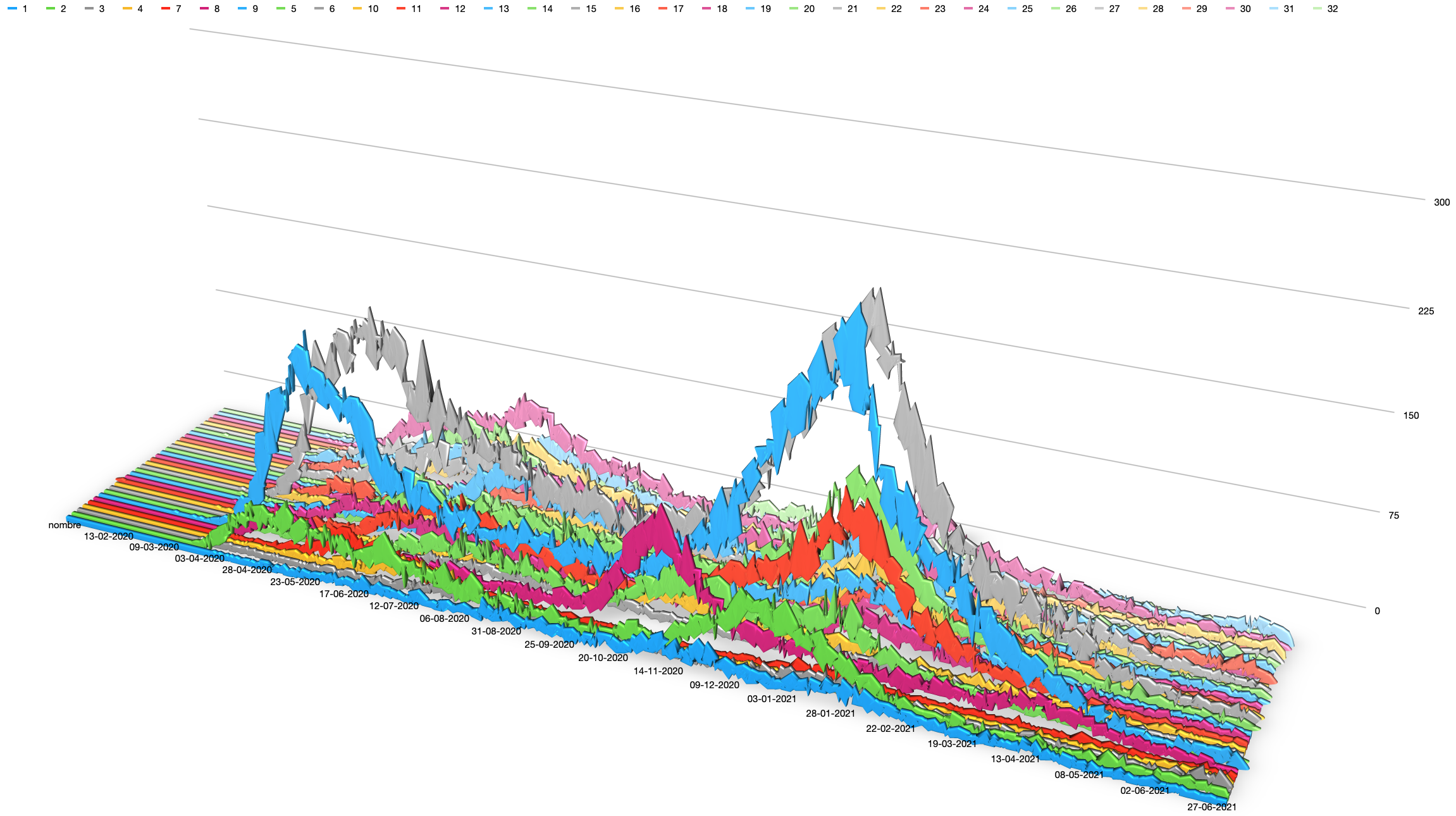}}} %\hspace{0.7cm}
\end{center}
\caption{Daily deaths register in Mexico for 32 states from February 2020 to June 2021. (a) Maximum deaths versus minimum deaths and (b) Total population versus deaths.}
\label{3DPlotDead32States}
\end{figure}

Figure~\ref{2DPlotDead32States} shows a tracing about deaths by COVID-19 in Mexico from February 2020 to February  2021. We calculate our own graph with data offered by the Mexican government.\footnote{Data COVID-19 M\'exico. \url{https://datos.covid-19.conacyt.mx/}} This graph shows the history of deaths every day for each of the 32 states in the Mexican republic. Here, we can see quickly which two main states are the epicenter in Mexico, they are: Mexico City (color brown) and the State of Mexico (color cream). It is not a surprise that the most populated city in Mexico has the main problem. Also, we must consider additional degrees of liberty such as: poverty, poor public health service, slow education, and local violence.

On the other hand, the Mexican culture has a particular point of view to understand the death from prehispanic times and actually with different type of violence linked to narcotraffic, violence against women, and multiple homicides. This way, the Mexican population is accustomed to see deaths with violence every week and every month. Therefore many people in Mexico do not believe in the virus and the perception is that the number of deaths is not representative. Part of this scepticism is exemplified in Fig.~\ref{3DPlotDead32States}a which displays a comparison between the total population versus the number of deaths by every state. This way, we cannot find maximums representative versus its total population, all values are below 1\%. Particularly, Mexico City has just a representation of 0.35\% whereas the State of Mexico has 0.20\% and for the rest of the states the probability is yet more slow.

We plot the same statistics about deaths by COVID-19 in Mexico but now in a three-dimensional representation in Fig.~\ref{3DPlotDead32States}b. In this graph, we can see the curves for the 32 states of Mexican republic as well, and clearly Mexico City (blue color) and the State of Mexico (gray color) concentrate the main problem, therefore it is the epicenter in Mexico. These correlations are close and it is because the population in Mexico City and the metropolitan area of the State of Mexico they are practically joined and transmissions between both entities is very high. So, the valleys in the graph illustrate that summer and autumn stations both were the best months with a reduced mortality in 2020, while winter displays the second wave of COVID-19. This second wave was the worst phase by the population in Mexico. Moreover, spring of 2021 displays the best moment in Mexico, showing a significant reduction of deaths and reaching more short maximums. Despite the fact that in Mexico several people do not follow the recommendations including the same government the summer seems encouraging. While some people cannot follow the recommendations by economic necessity, many others simply do not want to do anything.

We will note and emphasize that the Mexican government never implemented a strict quarantine and the people and several activities never stopped to the date. Indeed, the Mexican president (Andr\'es Manuel L\'opez Obrador) gives a very bad example as the people will protect themselves. In the same direction, the health minister (Hugo L\'opez Gatell) gives an appalling management of the pandemic, recommending safe distance, stay at home, the use of face mask but he made exactly opposite, and both persons are responsible for a lot of number of deaths in Mexico (for an ample number of details see \cite{ximenez2021criminal}).

%%%%%%%%%%%%%%
\section{Locality and globality}

\begin{figure}%[th]
\begin{center}
\subfigure[]{\scalebox{0.25}{\includegraphics{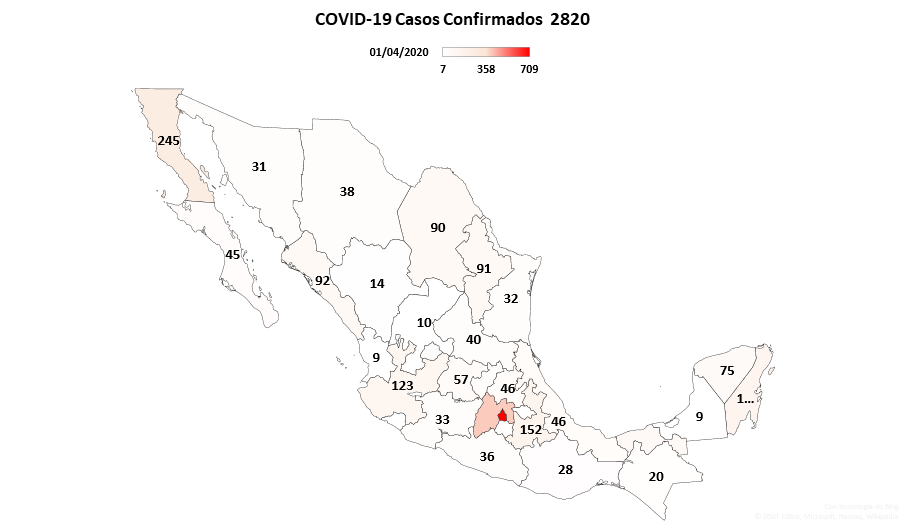}}} %\hspace{0.7cm}
\subfigure[]{\scalebox{0.25}{\includegraphics{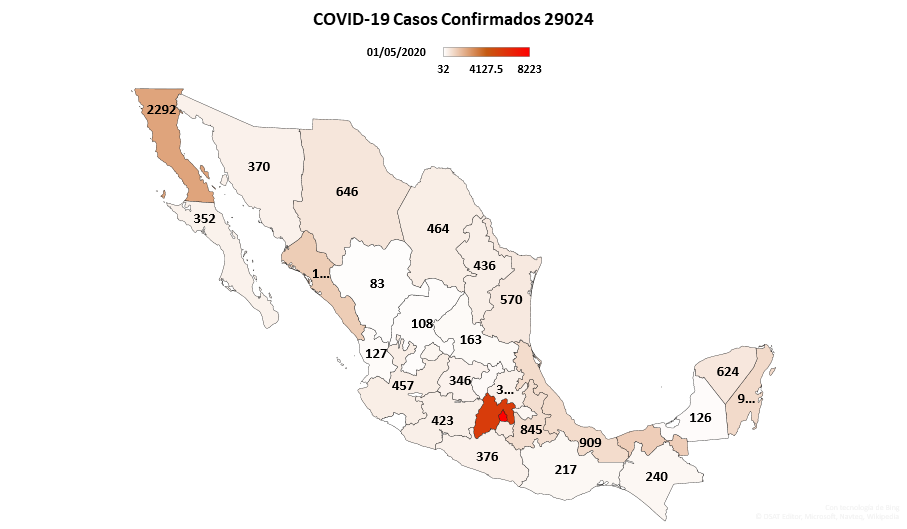}}} %\hspace{0.7cm}
\subfigure[]{\scalebox{0.25}{\includegraphics{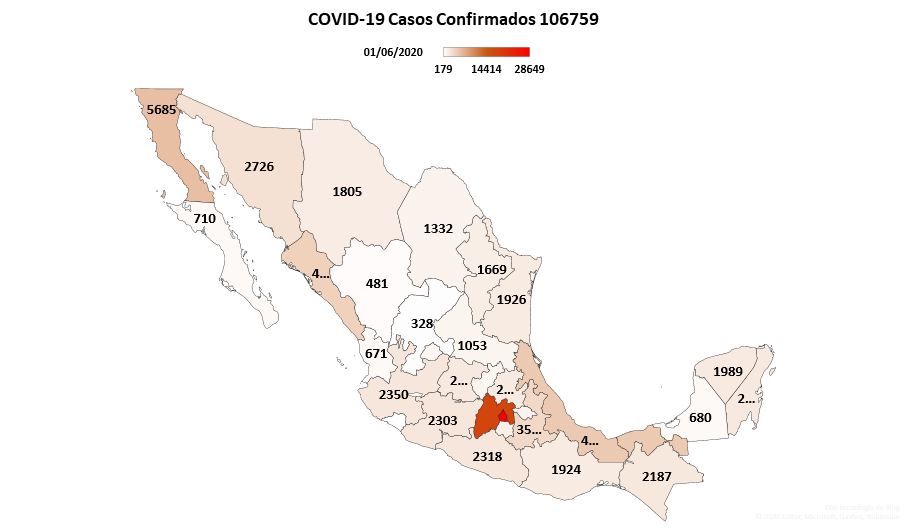}}} %\hspace{0.7cm}
\subfigure[]{\scalebox{0.25}{\includegraphics{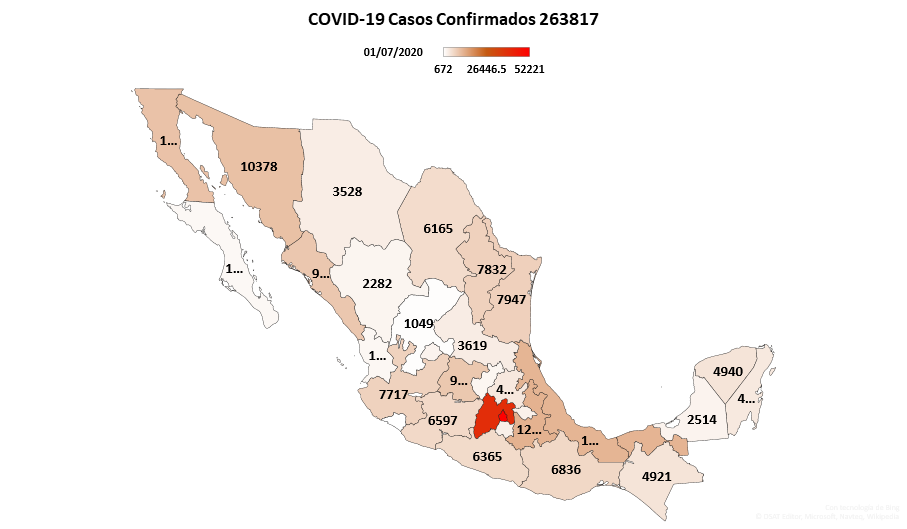}}} %\hspace{0.7cm}
\subfigure[]{\scalebox{0.25}{\includegraphics{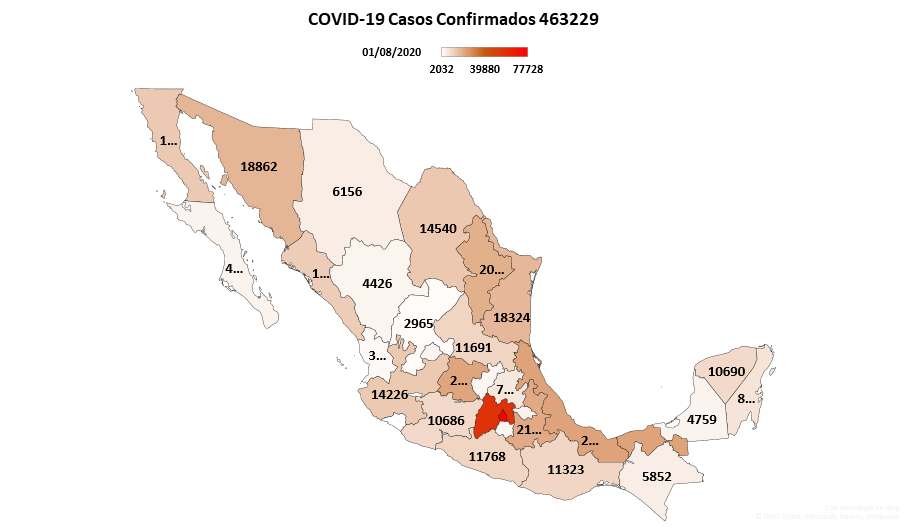}}} %\hspace{0.7cm}
\subfigure[]{\scalebox{0.25}{\includegraphics{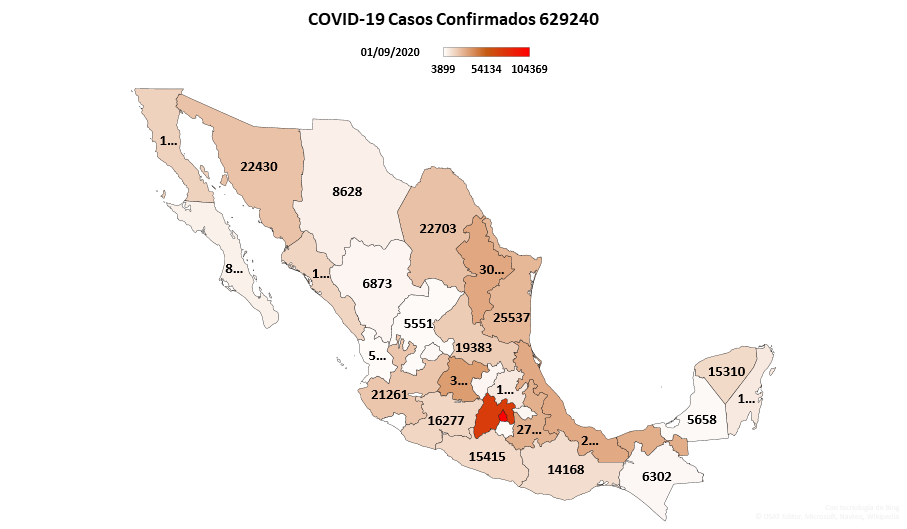}}} %\hspace{0.7cm}
\end{center}
\caption{Dynamics evolution of COVID-19 in Mexico. We show in every snapshot the first day per month in the year 2020: (a) April, (b) May, (c) June, (d) July, (e) August, (f) September.}
\label{maps1}
\end{figure}

\begin{figure}%[th]
\begin{center}
\subfigure[]{\scalebox{0.25}{\includegraphics{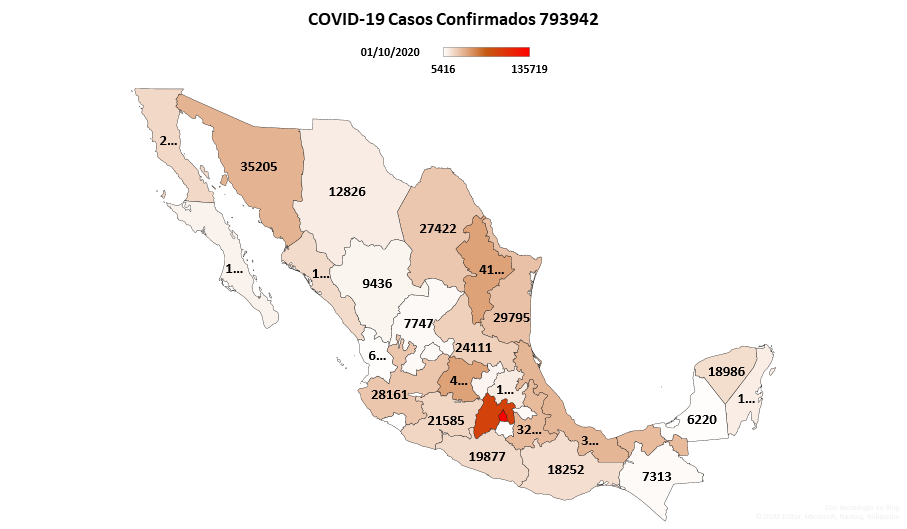}}} %\hspace{0.1cm}
\subfigure[]{\scalebox{0.25}{\includegraphics{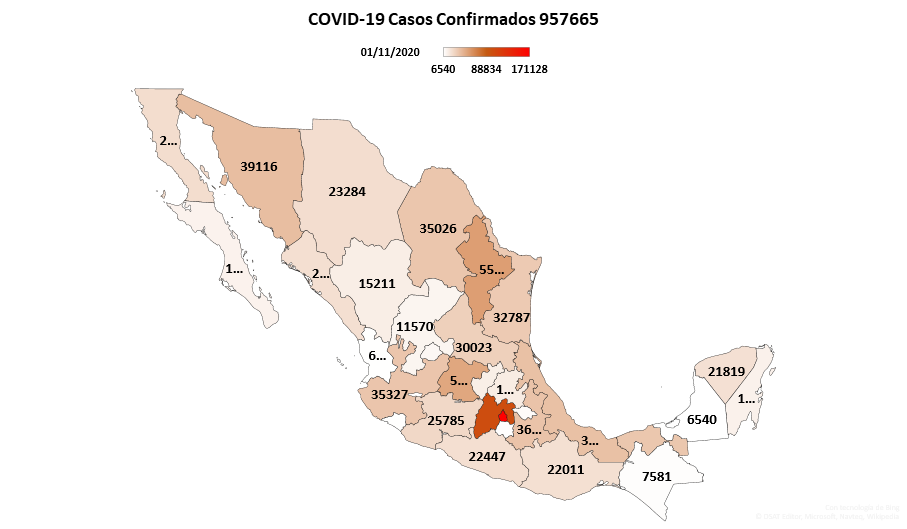}}} %\hspace{0.7cm}
\subfigure[]{\scalebox{0.25}{\includegraphics{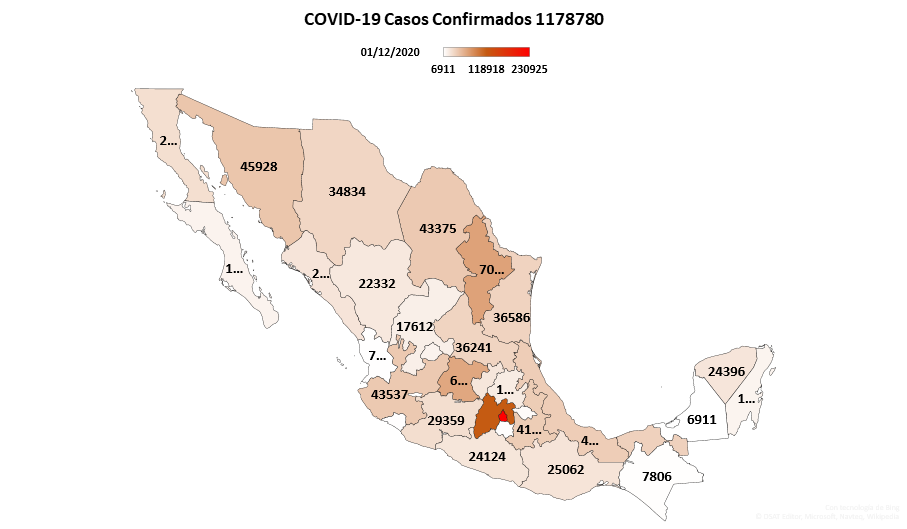}}} %\hspace{0.7cm}
\subfigure[]{\scalebox{0.25}{\includegraphics{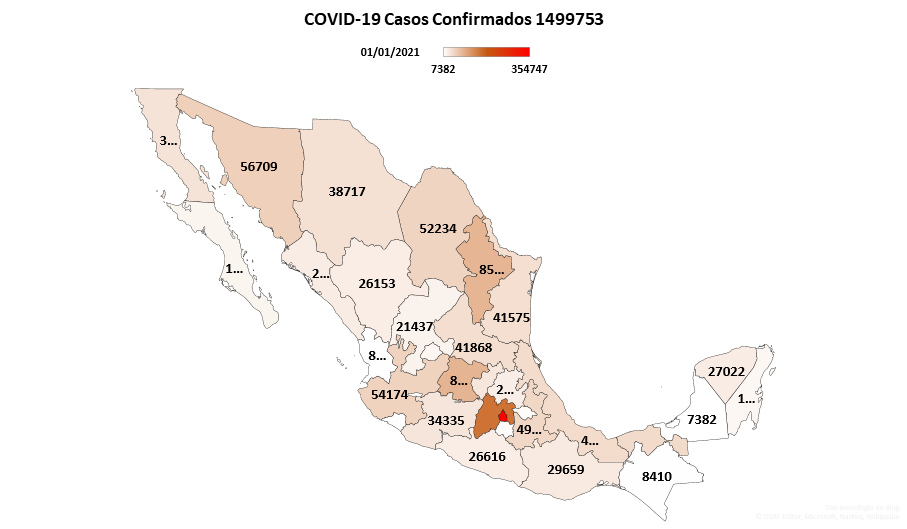}}} %\hspace{0.7cm}
\subfigure[]{\scalebox{0.25}{\includegraphics{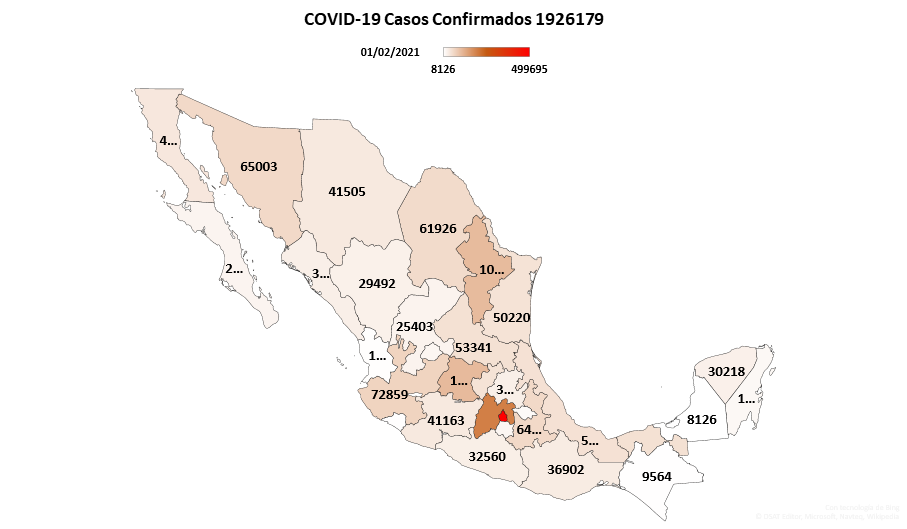}}} %\hspace{0.7cm}
\subfigure[]{\scalebox{0.25}{\includegraphics{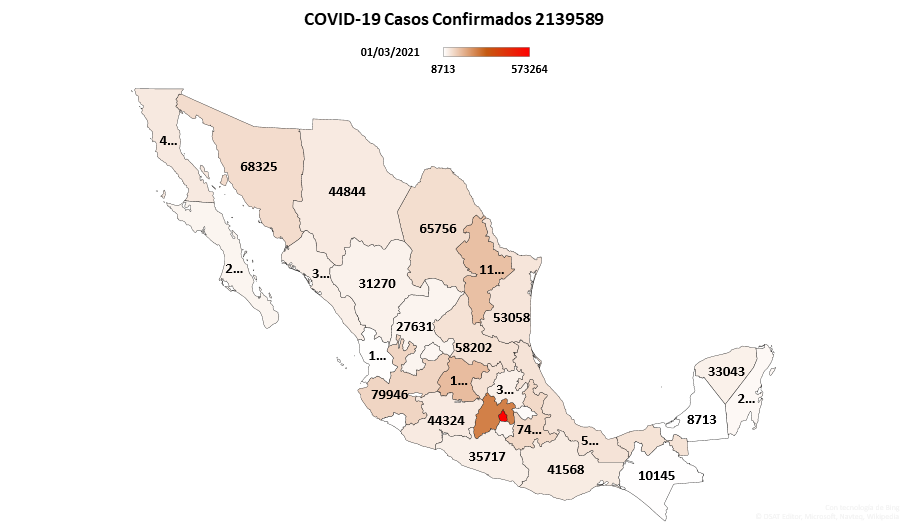}}} %\hspace{0.7cm}
\end{center}
\caption{Dynamics evolution of COVID-19 in Mexico months: (a) October, (b) November, (c) December, and (d) January, (e) February, and (f) March, in the year 2021.}
\label{maps2}
\end{figure}

\begin{figure}[th]
\begin{center}
\subfigure[]{\scalebox{0.25}{\includegraphics{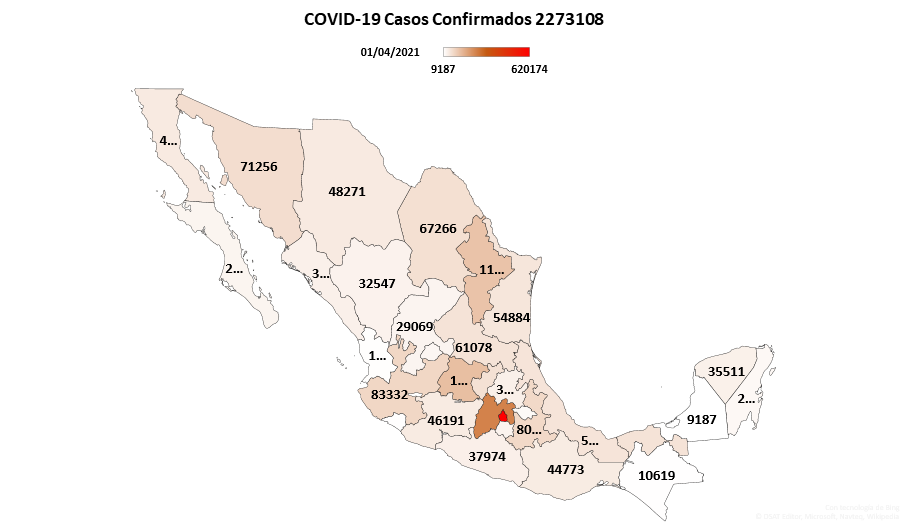}}} %\hspace{0.1cm}
\subfigure[]{\scalebox{0.25}{\includegraphics{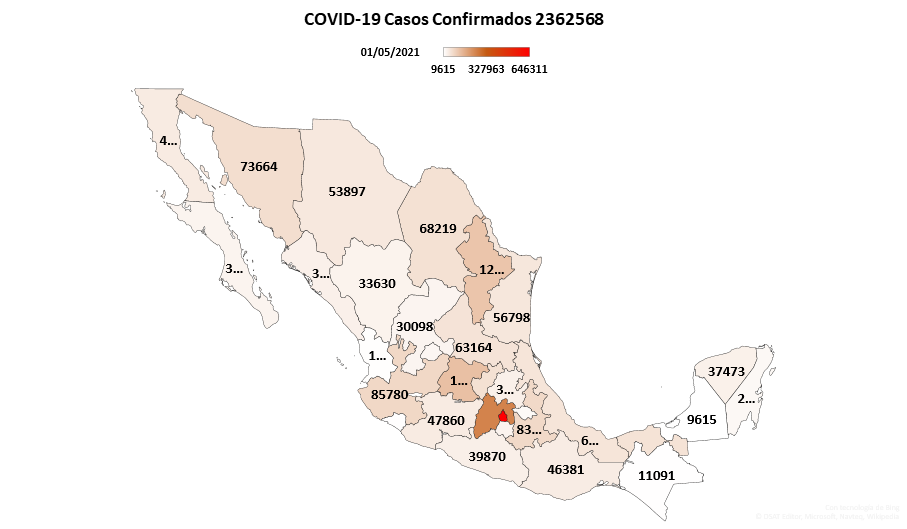}}} %\hspace{0.7cm}
\subfigure[]{\scalebox{0.25}{\includegraphics{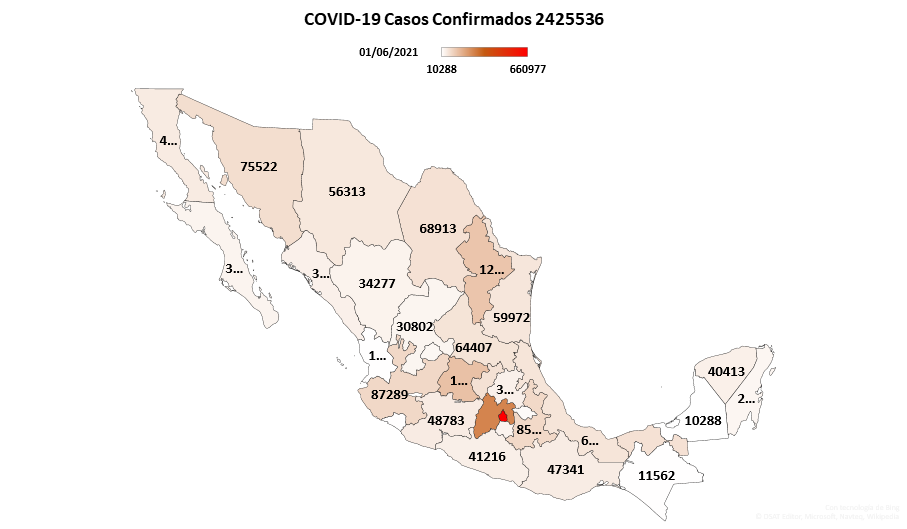}}} %\hspace{0.7cm}
\end{center}
\caption{Dynamics evolution of COVID-19 in Mexico months: (a) April, (b) May, and (c) June, in the year 2021.}
\label{maps3}
\end{figure}

As we show in the previous section, Mexico City and the State of Mexico are the epicenter of COVID-19 in Mexico. In this section we do a spatial analysis using maps of the Mexican republic. We use a range of color degradation between white and red colors. If a state is in color white hence the percentage with respect to the total number of positive cases of COVID-19 in this region is slow, while a red color represents the maximum value in this state and therefore a critical point in those dates.

In Figs.~\ref{maps1} and~\ref{maps2}, we show the history of positive cases registered in the Mexican republic per month during one year. We start in April because really in this month the people and the government started to display a real worry when the virus could be detected from January or from December 2019. Every map displays the first day of each month. So, the main characteristic of these maps is that colors are all adjusted with respect to the maximum value thus the main epicenter is preserved although the number of cases increases. In each map we can see numbers into every state. It is the number of positive cases, for the first months these numbers can be read but later it is more complicated to put the full number. On top every map displays the minimum value of positive COVID-19 cases registered (white color, left side) to show the maximum value (red color, right side). This way, Fig.~\ref{maps1}a says that the minimum number of positive cases registered in the Mexican republic to April 1st was 7 positive cases registered in Tabasco while the critical point was Mexico City with 709 positive cases registered. But for February 2021, Fig.~\ref{maps2}e shows us that the minimum cumulative number is 8,126 in Campeche while the critical point registered is nearly half a million of persons in Mexico City. From April 2020 to February 2021 we can see that the epicenter in Mexico does not change. Particularly some states showed subcritical densities about positive cases, such as: Baja California Norte in May and June (Fig.~\ref{maps1}bc), Sonora and Veracruz in July (Fig.~\ref{maps1}d), Puebla in August (Fig.~\ref{maps1}e), Nuevo Le\'on, and Quer\'etaro in September (Fig.~\ref{maps1}f), October, November, December, January and February (Fig.~\ref{maps3}abcde). Therefore the subcritical region most affected by COVID-19 in Mexico is the center-north with Nuevo Le\'on, and Quer\'etaro.

\begin{figure}[th]
\centering
\includegraphics[width=0.6\textwidth]{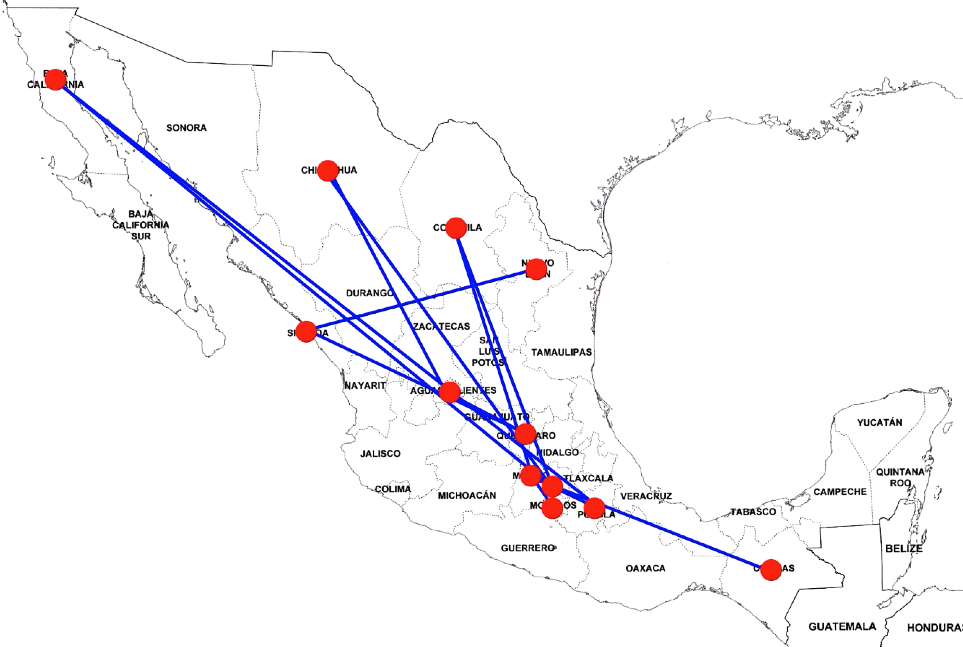}
\caption{Maximal deaths per state in Mexico during one year, from April 2020 to February 2021.}
\label{CriticsPoints2}
\end{figure}

In general, by regions the center-north of Mexican republic is the most affected by COVID-19 highlighting of course the center of Mexico. It is consistent if we return to check the maximum cases of deaths per state. In this sense, Fig.~\ref{CriticsPoints2} shows the maximal number of deaths registered one day everyday. These critical points registered each day are shown from February 2020 to March 2021. To this date the states registered with critical points are: Mexico City, State of Mexico, Morelos, Puebla, Chiapas, Quer\'etaro, Aguascalientes, Sinaloa, Nuevo Le\'on, Coahuila, Chihuahua, Baja California. Which implies the 37.5\% of the Mexican republic. Again, the most high frequency is dominated fully by Mexico City and the State of Mexico.

The same analysis is done locally for Mexico City. In this case, the main epicenter emerged in Iztapalapa, Gustavo A. Madero and Tlalpan. For the State of Mexico the epicenter emerged in Ecatepec and Neza City. Immediate correlations exist between:

\begin{center}
Gustavo A. Madero $\longleftrightarrow$ Ecatepec, and \\ Neza City $\longleftrightarrow$ Iztapalapa. \\
\end{center}

In this stage, we can conclude for the Mexican case, that cities with a high density of people implies a good opportunity for the virus and it is strengthened under very complicated conditions of poverty, unsafety, and very poor health services.

\begin{figure}[th]
\centering
\includegraphics[width=1\textwidth]{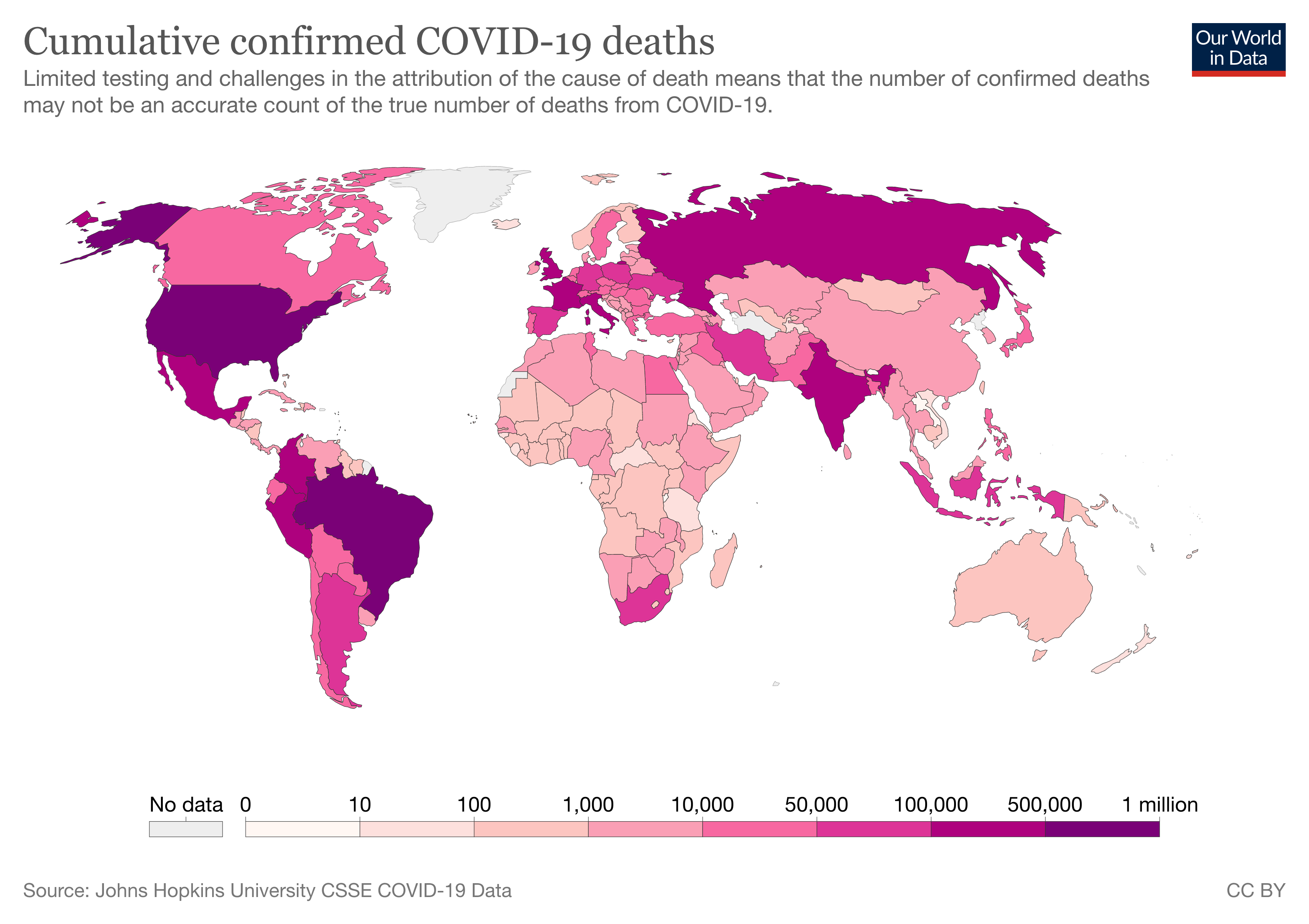}
\caption{Cumulative deaths around the world (this plot is calculated from the site “Our World in Data”). So, following the same analysis, globally we can see that the epicenters around the world are in America: USA, Mexico and Brazil. In Europe: UK. In Asia: India. In Africa: South Africa. In Oceania: none.}
\label{WolframAlpha}
\end{figure}

Following this criteria (number of deaths) for the whole world, we can see that the global epicenters are: USA, M\'exico and Brazil (in American continent), UK, France and Italy (in Europe), South Africa (in Africa) and Rusia, India and Pakistan (in Asia), while the Oceania continent no. The problem to understand the COVID-19 dynamics around the world is a complex system and it is clear that actually it is impossible to predict. Therefore, in many countries they could have handled the pandemic situation in different degrees of levels and procedures. Consequently, we can see that not all countries around the world (cities, towns) would be punished economically, socially, educationally and psychologically.

%%%%%%%%%%%%%%
\section{COVID-19 dynamics as a complex system}

\begin{figure}%[th]
\centering
\includegraphics[width=1\textwidth]{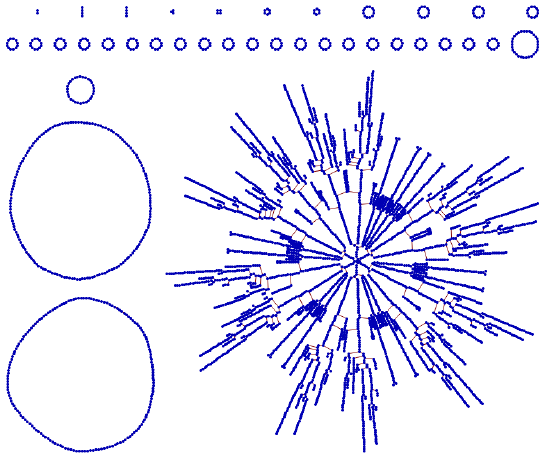}
\caption{A complex system shows the slow probability where an outbreak could arise. In this case, a binary discrete dynamical system as a one-dimensional cellular automaton classified as complex, evolving in a space of $2^{12}$ configurations. Moreover, the number of connections where a configuration could evolve with just one ancestor is determined in 36 cycles. This way, the probability where a field of configurations could have multiple ancestors and a most high propagation of them is $1/37$ that is a $2\%$ in this population.}
\label{ECA106ring12}
\end{figure}

In \cite{r02020Moore}, Cris Moore determines that the pandemic dynamics is not the same in any country, showing two examples where an average with a probability and ratio of propagation does not always follow the criteria for controlled systems. Moore exposes how the rate of transmission is an average of $R_0$ and the number of outbreaks can vary yet in subcritical conditions. He exposes how this propagation is related to the number of infections starting with a sickness, i.e, the number of infections starting from a person. This way, if $R_0 < 1$ then the number of infections and propagation is low but while $R_0 > 1$ then there exists a major probability to get a high number of infections.

In this paper, we compare this perspective in some well-known complex systems in the domain of one-dimensional cellular automata. This kind of discrete dynamical systems are very useful to study: the nature of complex systems, artificial life, computability, reversibility, mathematics, physics, biology, sociology, languages, and beyond \cite{mcintosh2009one, adamatzky2018cellular, griffeath2003new}. 

\begin{figure}[th]
\centering
\includegraphics[width=1\textwidth]{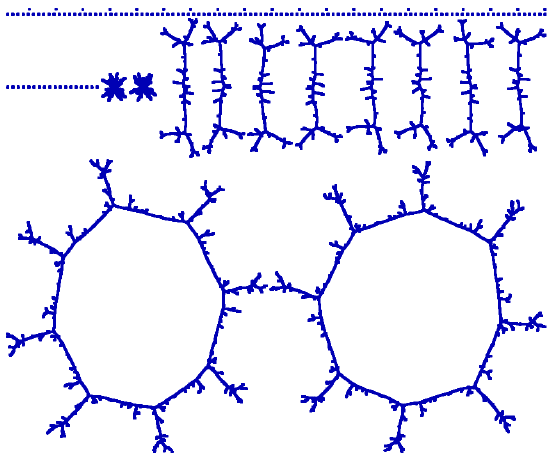}
\caption{In this case, we have the same complex system but now running on a space of $2^{16}$ configurations. In this case, the complex system is able to produce 8,224 different subsystems where only four of them are not equivalent. This way the probability to get an outbreak with a high number of ancestors is 0.1\%.}
\label{ECA106ring16}
\end{figure}

A complex system is a system conformed by primitive elements and their interactions conduce to components with emergent behaviour, non-trivial collective behaviour, self-organization, adaptability, criticality, evolution, and unpredictability  \cite{mitchell2009complexity, mainzer2011universe}.

A classic problem in cellular automata theory is the recognition and classification of complex functions. We consider  well-known one-dimensional cellular automata functions with complex behaviour and we exemplify this global behaviour across circle diagrams (attractors) and they represent by plots the evolution diagrams. This global relation has demonstrated in several previous researches in \cite{wuensche1999classifying, martinez2013designing, martinez2018simple}, that the complex behaviour displays:

\begin{center}
{\it long transients, moderate densities, non-symmetric ramifications, and moderate elements in the attractor}.
\end{center}

Particularly, we consider a complex one-dimensional cellular automaton and calculate two domains of their cycle diagrams to expose this phenomenon. It is the elementary function rule 106 proof complex by composition of memory functions \cite{martinez2013designing}. Particularly, rule 106 is a resistive rule able to preserve its dynamics later of compose its original function. Thus Fig.~\ref{ECA106ring12} and Fig.~\ref{ECA106ring16} show the dynamics of cycles diagrams for binary strings with lengths 12 and 16 respectively. Given a global configuration (node) of the system we apply the local rule to determine the next global evolution (transition). This way, we have an evolution space of $2^n$ configurations. Several of them represent cycles with a determined number of nodes (its period), and others determine trees, where we can see ancestors of configurations in its branches and leafs. In both cases, we can see that many configurations have a low probability to have multiple ancestors, because several of these dynamics reach a stability quickly or in a few steps. While few configurations have a probability to get multiples and a large number of ancestors. A dynamics explained and illustrated by Moore in his study \cite{r02020Moore} to COVID-19 dynamics.

\begin{figure}[th]
\centering
\includegraphics[width=1\textwidth]{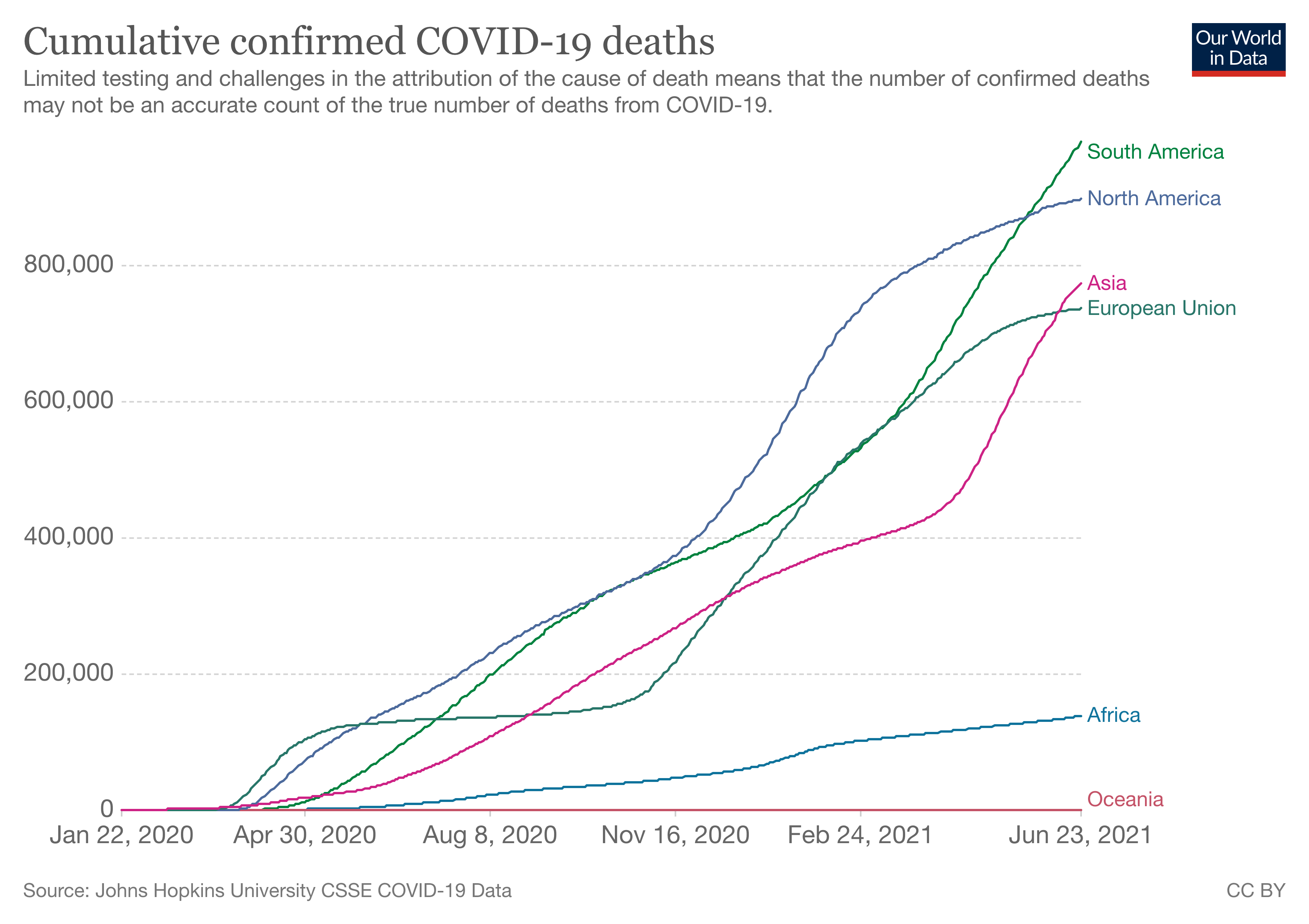}
\caption{This plot is calculated from the site ``Our World in Data''. Shows the relation by the number of cumulative deaths registered by continents around the world during one year.}
\label{coronavirus-deathsCom-June2021}
\end{figure}

COVID-19 dynamics around the world display several of these properties and it can be classified as a complex system. 

\begin{itemize}
\item In what country could emerge a new contagious and lethal COVID-19 mutation? It is non predecible and it is impossible to determines where could emerge a new danger mutation.
\item Following several densities curves by countries around the world all they are heterogeneous. It is the same phenomena by continents, cities, towns, i.e, locally and globally.
\end{itemize}

It is interesting as the critical points change geographically sequentially by continent. This way, we can see these epicenters moving in different stages successively, in the following transition:

\begin{center}
$Europe \rightarrow America \rightarrow Asia$ (see Fig.~\ref{coronavirus-deathsCom-June2021}\footnote{Our World in Data. \url{https://ourworldindata.org/coronavirus}}).
\end{center}

%%%%%%%%%%%%%%
\section{Final remarks}
Reviewing and following data about COVID-19 in Mexico, we can recognize the critical regions with more affections by this virus in Mexico. This way, globally Mexico City and the metropolitan area are the critical point in the Mexican Republic. They concentrate the most high density of population. While locally, Iztapalapa, Gustavo a Madero and Tlalpan were the critical points in Mexico City.

We have discussed in the section 3 which the country could have handled with different restrictions and not homogenize a social and economical punishment with the same level. A real epidemiological vigilance in all the country was completely ignored by the Mexican government.

Finally, we showed as the COVID-19 dynamics is related to a complex system dynamics.

%%%%%%%%%%%%%%

\end{document}